\documentclass{article}
\usepackage{hyperref}

\newcommand{\email}[1]{\texttt{\small #1}}

\begin{document}

%%
%% Title etc.
%%
\title{Inheritance and Blockchain: \\
      Thoughts and Open Questions }

\author{Fr{\'e}d{\'e}ric Prost \\
        LIG - CNRS and Univ. Grenoble Alpes, Grenoble, France \\
         \email{frederic.prost@univ-grenoble-alpes.fr}} 
\maketitle

%%%%%%%%%%%%%%%%%%%%%%%%%%%%%%%%%%%%%%%%%%%%%%%%%%%%%%%%%%%%%%%%%%%%%%%%%%%%%%%
%%%%%%%%%%%%%%%%%%%%%%%%%%%%%%%%%%%%%%%%%%%%%%%%%%%%%%%%%%%%%%%%%%%%%%%%%%%%%%%
\begin{abstract}
Inheritance is the fundamental building block of civilization. This is the
addition of wealth, knowledge and properties over time that produce the society
in which we are living. Every generation does not have to start from zero and
can capitalize on the efforts of previous generations. Blockchain based assets
are very efficiently and securely transferred between living entities. Yet the
actual way to make heirs inherit crypto-assets is seldom discussed. It appears
that the problems linked with the inheritance of crypto-assets raise a lot of
technical, societal and legal issues. Part of those issues have to be tackled
with at the level of the blockchain infrastructure itself. The aim of this paper
is to open a research field, and to discuss some ideas, with regards to this
overlooked issue. Inheritance is neither a peripheral question nor one that can
be dodged.  It comes with its own set of challenges that have to be met if
blockchain based finance, and asset management, is to be taken seriously.
\end{abstract}

%%%%%%%%%%%%%%%%%%%%%%%%%%%%%%%%%%%%%%%%%%%%%%%%%%%%%%%%%%%%%%%%%%%%%%%%%%%%%%%
%%%%%%%%%%%%%%%%%%%%%%%%%%%%%%%%%%%%%%%%%%%%%%%%%%%%%%%%%%%%%%%%%%%%%%%%%%%%%%%
%% main text
\section{Introduction}
\label{sec:introduction}

Crypto-currencies, starting with Bitcoin \cite{Nakamoto_bitcoin}, have shaken
the world of finance in less than a decade. It has moved from a pipe dream
concept to an every day tangible reality in the meantime. At the time of 
writing the global market capitalization of emitted bitcoins is around 10\% of 
the global market capitalization of gold. There are many discussions on the 
nature of money, and assessing the relative merits of Bitcoin versus gold as a 
store of value is still an ongoing discussion.

Among the mandatory properties that a store of value must have, the 
property of being inheritable is a major one. Over a sufficiently long period the survival rate of
everyone drops to zero. Transmitting wealth to the future generations is neither a
peripheral issue, nor one that can be dodged. The body of laws, stories and
traditions about inheritance is immense. In fundamental texts like the Bible
\cite{Bible} or the Odyssey \cite{Odyssey}, the question of who inherits what
from whom, and more generally all kinds of problems linked with succession, are
major preoccupations. 

The issue of inheritance is orthogonal to the actual implementation of the store
of value. Society, in a very broad sense, is the tool traditionally used to
transfer titles, assets, and to settle questions like: "Who is the new King?".
Regarding material wealth, objects do not disappear when their owner dies. These
remarks no longer hold with crypto-assets. Indeed, one fundamental feature of
the crypto-assets is that it is only with the knowledge of the appropriate keys
that the assets can be transferred. The actual ownership of crypto-assets amounts
to the knowledge of the keys and vice-versa: everyone that knows the keys is
deemed to be a rightful owner of the associated crypto-assets. But by
definition, and under these circumstances, one cannot actually implements his/her
own succession because either the knowledge of the keys disappears with their
death or the ownership of those crypto-assets has to be shared with a third party. 
It appears that we are finding ourselves painted into a corner: on one hand a 
crypto-asset should only be transferred by its legitimate owner; and on the other 
hand, the owner cannot transfer anything once dead, making the succession of 
crypto-assets seemingly impossible.  

In this paper we discuss various challenges and ideas linked with the
inheritance of crypto-assets from a computer science point of view. The idea of
the discussion is to think about what desirable properties should be looked for,
in the original spirit of crypto-currencies, inheritance of crypto-assets.
Indeed, the easiest way to "solve" the inheritance problem would be to write the
keys into the wills. This solution is going to be discussed further later in the
paper in section \ref{subsec:delv_compl}. But clearly, it is not in the spirit of
crypto-currencies because it relies on a third party. It clashes with the
fundamental idea of crypto-currencies which is to be able to establish trust
without having to rely on third parties.  

In section \ref{sec:short_rev}, basic issues linked with the issue
of crypto-asset inheritance are discussed. Problems are addressed in a colloquial
way and presented by increasing level of complexity. In section \ref{sec:lit_rev} 
is done a state of the art review. Future themes for research and reflection are 
pointed out in section \ref{sec:themes}. We finally conclude in section
\ref{sec:conclusion}

\section{Informal discussion on the inheritance of crypto-assets}
\label{sec:short_rev}

   \subsection{Delving into the complexity}
   \label{subsec:delv_compl}
Let's examine some issues, as well as some workarounds, raised by the issue
of the inheritance of crypto-assets step by step. In this section we explore
this question from a naive point of view in order to give an intuitive idea of 
the landscape. In this section we focus on intuitive understanding of the 
issues, a more technical/disciplinary point of view approach is considered in 
section  \ref{sec:themes}.  

A one liner frame of the question to be addressed goes something like this: 

\begin{center}
   \textit{How can my eight years old daughter inherits my 
crypto-assets?} 
\end{center}

This is a starting point, there are many more subtler sub-problems. Actually, the
more the problem of inheritance is considered seriously, the more its inherent 
complexities appear. Let's examine a sample of those issues, together with 
some tentative solutions, by increased level of sophistication. 

   \begin{enumerate}
	\item The first idea to solve the basic ''eight years old daughter
inheritance'' problem is:
        \begin{itemize} 
           \item[(a)] To set up a meeting with a lawyer. 
           \item[(b)] To write down the wills on a document, including the 
                      appropriate private keys. 
           \item[(c)] To seal off the envelope.  
           \item[(d)] To hope for the best. 
        \end{itemize}

 This natural solution presents many challenges. The more salient being that
crypto-currencies have been built precisely to provide agreements without having
to rely on identified third parties.
There is maybe nothing as opposed to this aim than having to go to see a lawyer,
and having to rely on the professional integrity and competence of this lawyer.
This is a poster child of all the issues linked with centralization. There are
many additional issues. One can think of anonymity related issues: for instance
in order to establish the wills, a comprehensive list of all crypto wallets has 
to be done. Another class of problems is the risks inherent
to this solution. Typically for lawyers that will be targeted by wrongdoers
if this practice becomes mainstream (because stealing the keys give effective
control of the crypto-assets). The saying "not your keys not your coins" sums it
all accurately. The simple fact that you have to give your key to a third party
amounts to lose the ownership of the associated crypto-assets.

Essentially this solution reintroduces the single point of failure with all the
drawbacks attached to it.   

	\item The second idea that may come to mind is to put all the keys on a
thumb drive, or write them down on a piece of paper, and lock them into a safe at
home. It marginally improves on the previous point if family is more trusted
than professional lawyers. It is a more distributed solution because each
individual implements a particular version of this protocol. 

 Besides the hazards that such a practice would produce if it were widely
adopted (basically the same issue than the one discussed with the lawyer's
solution and the incentive for wrongdoers to steal keys), it has the following
additional drawback: actually my eight years old daughter is not my only heir.
Let's say there are four kids and seven nephews between whom the inheritance is
to be shared. It is not as if dramas about succession, struggles within
families, and communities, are a literary genre unto themselves.  Moreover, how
can someone be sure that the individual opening the safe will behave correctly?
It is harder to cheat with a pile of physical gold because there may be
witnesses, the material has to be physically moved around etc. With
crypto-assets, one just has to remember a passphrase. No one can stop someone
knowing the correct keys from using them later. Short of killing, it is
impossible to delete a passphrase from the memory of another person. 

	\item It is possible to be smarter and to write a smart contract that
implements the succession wills. It solves the "four kids, seven nephews"
problem: the smart contract is going to perform the sharing. It can be seen as
an equivalent of the wills in the blockchain world. It raises a new problem
tough: how will the blockchain be aware of the death of the owner of the smart 
contract? This is a variant on the famous oracle problem \cite{Oracle}. The
first ideas to deal with this issue would be to rely on some sort of state based
service reintroducing a centralization problem. This specific issue is discussed
more thoroughly in section \ref{subsec:distr_ad}. 
 
At this point another class of issues, that are not primarily technical but have
technical implications, appears: the heirs may not be of age to understand the
crypto technologies involved. Maybe the don't  have the legal rights to access such kind
of funds either. Some of the heirs may also not have crypto-wallets in the first place.
If so the mechanisms by which the proper credentials could be transmitted to 
them, without being compromised, remain mysterious. 

There is another type of issue: what if the four kids and the seven nephews die
with the one that they are supposed to inherit from? Let's say, for instance,
that they all (or any subset) disappear simultaneously in a plane crash? It is
not possible to re-write the smart contract: that is the very idea of smart
contracts. Then the inheritance disappears (more precisely it becomes
inaccessible) in such a scenario. In real life there are specific laws and legal
practices to deal with such kind of situation. This issue is discussed more
precisely in section \ref{subsec:flex_w} 

       \item A rather simple solution is to set up an equivalent of a time
capsule. If a date is chosen sufficiently far away in the future, then the
death of the capsule owner becomes a certain event. It can be done via smart
contracts that just have to wait until some block number is reached in the
blockchain before being executed. The drawbacks lie in the lack of flexibility
and the necessary approximation of the time of death. A middle-aged victim of a
traffic accident or an unexpected death could potentially lead to an inheritance
process stalled for more than half a century. Moreover, the probability that the 
potential beneficiaries of the inheritance may have died too in the meantime 
increases. 

       \item An improvement over the previous idea is to use a dead man's
switch. Instead of using the maximum age plus a safety margin for the time
capsule deadline, it is possible to use a shorter frame. If necessary one has
just to edit the time capsule deadline before it is executed. The
death of the time capsule owner stops this process of reprogramming, and the
time capsule is eventually delivered. It, partially, solves the issue of the lag
between death and succession. On the other hand it requires a constant vigilance
and work. It also opens possibilities of denial of service attacks. The denial 
of service can be malicious or due to life circumstances, typically the owner of
the dead man's switch cannot access the blockchain for technical or medical
reasons (for instance being in a coma etc.).  

	\item Everyone is going to die but we hope that it will happen as far
away in the future as possible. Life expectancy has improved a lot lately. From
a technical point of view it is a very challenging aspect of crypto-asset
inheritance to manage. It is very difficult to anticipate the technological
environment a few decades from now. However, any credible proposal for the
inheritance of crypto-assets must be resistant to the future. It suggests that 
any solution should be integrated within the crypto platform itself rather 
than having to rely on outsourced processes. This problem is more thoroughly 
discussed in section \ref{subsec:long_tp}.    
\end{enumerate} 

  \subsection{Decentralized Society and inheritance}
  \label{subsec:}

It is not yet clear how much blockchain based technologies are going to have a
fundamental impact on society. Bitcoin has changed the financial landscape in
deep. The most important central banks are considering to either develop their
own crypto-currencies and on ways to regulate this new field \cite{imf2022}.
Some actors go as far as suggesting that the nation-state framework could be
impacted \cite{networkstate}, or more modestly that a new kind of society, the
decentralized society \cite{Weyl2022} will emerge.

What is going to happen is very hard to discern precisely but it is clear that
the impact are not going to be restricted to the technical/technological tiers
of the society. The question of inheritance of crypto assets is going to be
major concern. 

There are already tricky issues linked with death and social
media: what is supposed to happen when the user of a social account dies? Is the
account frozen for ever? If people have access to this account (for instance
because the password was saved in a personal computer) can they use it? Should
the account be deleted? Those questions are not light and not easy to deal with.
There are propositions to use Artificial Intelligence in order to have a digital
clone of the dead person \cite{Cebo2021}. Is it ethical or not to impersonate a
dead person through an artificial intelligence piece of software? On what
grounds? Likewise there were studies on security implications of social media
account of deceased persons \cite{Dickerson-Southworth22}. Those accounts
contain many important data that can be used before the knowledge of the death
has been spread. 

It turns out that those questions, though difficult and important, are just a
small aspect of a larger one: what becomes the concept of legacy in the digital 
space? The problem of the inheritance of crypto-assets is a part of this larger 
issue. It raises challenges that are both technical, technological and societal.      

\section{Existing solutions review}
   \label{sec:lit_rev}

At the time of writing there has been very few proposals to tackle with the
various issues raised by the inheritance of crypto-assets. Some workarounds 
have been proposed. 

\begin{itemize}
   \item Sarcophagus \cite{sarcophagus} is a dead man switch implementation that
is blockchain-enabled. It is resistant to censorship and immutable. It is done
by the combination of Arweave \cite{Williams2017ArchainAO} for a permanent
storage of data, and Ethereum \cite{Buterin2013} to support the ERC20 Sarco
Token. This token is used to pay so called "archaeologists" which are in charge
of releasing the data (essentially an encrypted file) to the person of interest.
The user have to select one or more existing archaeologists. The archaeologist
public key is used as an outer layer of encryption. This outer layer has to be
re-wrapped at predefined dates in the future. If one date expires then the
archaeologist decrypts the outer layer. The inner layer is the data encrypted
with the public key of the final receiver that can decrypt it.    

In addition to the problems linked with the dead man switch that are  discussed
in section \ref{subsec:delv_compl}, there is the presupposition that the
receiver of the time capsule is already well identified and as access to the
necessary public key infrastructure.

   \item Ternoa \cite{Ternoa} is a french start-up that proposes a "death
protocol" which is basically a smart contract triggered by the API's of local
authorities registering deaths. It presents the problem of relying on a
centralized Oracle. One issue is that it is easier to hack the local authorities
database (or to bribe agents working for this agency) than to break a
distributed solution relying on crypto technologies. Another issue is that there
is no standard API to deal with this issue that is shared among countries.
Each solution is limited to one nation-state at best. Finally there is no
warranty that the API are not going to change in the future.   

    \item Casa \cite{Casa} is a company that proposes solutions based on
multi-signature schemes. Their primary service is to provide better resiliency
for crypto wallets. They also have an inheritance product that is basically a
technological implementation of the second point examined in section
\ref{subsec:delv_compl}. 
 
\end{itemize}

\section{Themes for future research}
\label{sec:themes}

   \subsection{A distributed protocol for the death announcement}
   \label{subsec:distr_ad}

   The aim is to define a protocol that is safe, distributed and has some
privacy properties such that the blockchain is aware of the death of a
particular individual. This is the basic signal that is going to be used as a
trigger for smart-contracts, whatever they might be (see \ref{subsec:flex_w}),
implementing the wills.   
   
   Every solution to this problem must at least meet the following criteria: 
   \begin{enumerate}
      \item It should be adaptable to any blockchain modulo  an appropriate
tuning of technical details and of governance peculiarities of the considered 
blockchain. 

      \item It should respect privacy in the sense that before the death has
been enacted by the blockchain, there should be no way to link specific wills,
whatever there form are, to a specific crypto-wallet/crypto-address. 

      \item It should present some warranties of a good execution, namely that
the inheritance will be done as it was planned. This point is not trivial because the
solution has to rely on a distributed system for which it is generally hard to
have hard warranties of execution.   

   \end{enumerate}

   In \cite{tales_from_crypt}, I have proposed such a protocol. The idea behind
the \emph{Tales From the Crypt Protocol} (TFCP) is the following: the signal of
the death is set when a predetermined amount of coins is transferred to a
special wallet after a predetermined time has elapsed. The coins are stacked by witnesses 
that testify on the death of a particular person. This stack can be lost if it
turns out that the information of death proved false. The proof of life is
adjudicated by the existence of a financial move on the account of interest. Once the network has
acknowledged the death signal, then another group of actors decrypt the
information allowing to make the link between the dead person and the
corresponding crypto-assets and smart-contracts. I refer to the paper
\cite{tales_from_crypt} for the technical details. 

   \subsection{Transmitting secrets to the future}
   \label{subsec:trans_sf}

Cryptography is very efficient at allowing multiple parties to exchange secrets.
The untold assumptions of this field are that both parties have to be \emph{alive}, and most of the
time \emph{identifiable}. Both of those assumptions may not be necessarily true
in the case of inheritance. One edge case is that the heir might still be in the
womb when the giver dies. There are also those stories of the research of the 
heirs that takes years or even decades. There is even a profession "probate 
genealogist" whose job is precisely to solve work on those kind of puzzles.    

 Without having to go as far as these extreme cases,  it is
clear that, within the realm of crypto-assets inheritance, the issue of 
transmitting the right credentials to the right person is a major preoccupation. 
Indeed the knowledge of
the keys amounts to the proof of ownership of those crypto-assets. Thus the
paradoxical requests: the secrets have to be transferred to an unknown third party
at an unknown time, and in the meantime there should be some warranties that the 
secret is not revealed to any other party while the owner of the secret have
disappeared. The fact that the time lapse can very easily be counted in years
only makes the issue more complex. 

A first idea would be to adapt some secret sharing scheme \cite{beimel2011}, but
it doesn't look a trivial endeavor at first sight. Indeed, whatever the scheme
would be, the fact that shareholders should be forbid to collude to reveal the 
secret is harder to implement: as we mentioned time lapse is very large. In
facts it is so large that shareholders might have to transfer their shares to another
shareholder for instance. 

Another idea to develop is to objectify shares in order to regain some control.
The raw idea would be to find ways to produce clones of objects that do not
require any digital process. A very simple idea is the following: you can
produce material keys (pieces of metal) that are clones using key duplicating
machines. Those machines are analogical: there is no file recording how the
clone can be produced. Then these clones can be used in the future as a shared
secret: by making standard measurements on the object, it would be possible to
generate some bits of shared information. One interesting point is that the
measure has not to be fully specified at the time of production. The only
property that has to be met is that they are really clones from one another,
meaning that any physical measurement on both objects  will give the same
results. Another desirable property would be that it should be materially
difficult to produce other clones: simply scanning the object should not be
enough. The underlying idea is that those material keys are going to be saved in
a physical safe, and just having access to them should not be enough to make
clones. Another important property that such an artifact must have is the
stability of the object across long periods of time. The 
physicality/materiality aspect of this kind of solution  could be a way to work
 around the  issues raised by the non destructive, and perfect copy of 
information, that digital technologies allow.      

   \subsection{Flexibility of wills}
   \label{subsec:flex_w}

One of the very basic motivation behind the smart contracts \cite{smart2022} is
the fact that once they have been enacted, it is impossible to change them. This
feature is very interesting in a variety of application. In the case of
inheritance, typically when the wills are implemented via a smart-contract, it
can be a hard problem to solve. Indeed, as discussed above, the identity of the
heirs is not always the one that was initially defined. There are more subtle
problems, like, for example, some evolution of the inheritance laws between the
coding of the wills. Those problems essentially comes from
the fact that there is an incompressible part of the inheritance process that
cannot be completely foresaw in advance. Therefore, developing smart-contracts 
more flexible is an important task from this perspective. 

This has not to be an evolution of the fundamental ideas behind
smart-contracts but rather a way to develop an ecosystem that adds overlays
on top of the smart-contract in order to achieve more flexibility. Because death
is both unavoidable and unique it can be implemented as a set of ad hoc 
governance rules. Yet those rules have to be clearly determined and studied.
Their interactions with the rest of the blockchain could be tricky to precisely
analyze.   

   \subsection{Blockchain and civil status}
   \label{subsec:block_s}

The inheritance process brings to the forefront many problems around  links
between the virtual world and the material world. One of these issues is known
as Oracle's problem: how and by what rules is the blockchain "aware" of what is
happening in the outside world? Regarding the inheritance process there is a dual
type of issue: how is the blockchain being able to reach the outside world? If
you narrow the problem down to first principles, in order to perform an
inheritance one has to reach specific individuals. A first problem is that
people that are looked for do not necessarily have a presence in the blockchain
of interest. In the basic "8 years old daughter" scenario the problem is
illustrated by the fact that young children don't have a crypto-wallet. Another
similar issue is that, even if such crypto-wallets exist, they might not be
known by the giver. How can it be represented within the blockchain? There must
be some kind of link between the blockchain and the real world, i.e. the "social
security name" of heirs. Again you don't want to depend on third parties for
this link.

A recent proposal by Weyl, Ohlhaver and Buterin in \cite{Weyl2022} is the idea
of "Soul Bound Tokens" or SBDs. The idea is to have NFTs that can't be
transferred.  Their use could be to represent non-transferable and persistent
social relationships. Many legacy issues could use such a novel proposition in
order to bridge the gap between virtual and real world.

   \subsection{The long term problem}
   \label{subsec:long_tp}

Any solution to the inheritance issues of crypto-assets is going to face a
particular challenge: the solution is not going to be used before a, hopefully,
very large lapse of time has elapsed. Since it is not possible to foresee the
exact time of death, the inheritance process of young people may have to stay
put for many decades before being actually used. It is always possible
to rely on the end user to continually adapt to the new technological
environment. But this is clearly not a satisfactory answer both from a
practical, societal and philosophical point of view.      

From a practical point of view, the inheritance issue is not something that one
should have to work constantly on. It should be set once for all unless very
specific events occur: loss of an heir, new wedding in your family, apparition
of new heirs etc. Changing your wills when such extreme events happen is normal,
but otherwise one shouldn't have to tinker with his/her wills on a regular
basis. Remember than most of the time, in most civilizations, there are no
explicit wills. There is a "by default" mode, embodied into customs and
dedicated laws, that applies. Wills are used when something specific has to be
implemented, and more often than not laws restrict the extent to which wills can
be differ from the default procedure.  
  
Inheritance is the engine behind culture building. Therefore, the process of
inheritance cannot be completely let to the hand of individuals. In the same way
that there are basic laws (murder, stealing are forbidden etc.) for a society to
work, there are basic rules regarding inheritance that do not simply rest on
subjective/personal choices. This has strong implications on the inheritance of
crypto-assets, indeed any implementation might have to change because
legislation has evolved. On the course of many decades it is not surprising, or
even unexpected. This is somewhat contradictory with the previous point. The
fact that maybe there are parts of the inheritance process that are going to
evolve do not solely rest on the whim of individuals.  

From a more abstract point of view the inheritance process is by nature a
trade-off between individual wishes and societal rules. This trade-off slowly
evolves as time goes by. This has an impact on any technical "solution" to the
inheritance of crypto-assets. The whole idea of the blockchain governance must have
to take this into account. How are these issues going to be solved remain
subject to many trial and errors, but unlike most of other functionalities that
can be tested in real time, the inheritance functionality cannot be tested on a
large scale quickly. The development of simulations is both mandatory and
difficult: what are the correct model of users? How the fact that rules are
going to change over time can be coped with? Those are some of the questions 
that have to be tackled with.      

   \subsection{The customer is not the end user}
   \label{subsec:customer_eu}

Inheritance can be looked at both from the giver's side and from the heir's
side. What is unique in the "inheritance application", viewed as a feature of
the blockchain, is that the giver, by definition, may never check whether things
have unfolded as planned, and that the heirs may not even be aware that they are
heirs. Therefore the inheritance application is a very specific kind of
application for which incentives are very hard to set up correctly. Indeed,
almost by definition there are no customer feedback. Even if you inherit,
meaning that the processed somehow worked, how do you know that the process was
correctly executed? Moreover, because the inheritance may include privacy
management it is not even clear that the heirs of a single giver may know each
other. So they cannot regroup together to check that whole process has worked as
it was intended to. This is not intrinsic to the inheritance of
crypto-assets. But it is relatively new from a computer engineering perspective.
Indeed, in real life institutions, and society, are set to solve this issue.
Somehow the community "knows" who is who and what belongs to who, even if this
knowledge is distributed and somewhat fragmented. There are informal ways to
have feedback which is also known as "reputation". An equivalent has to be found
in the world of blockchain.   

Another related point is that heirs may refuse the inheritance. This is
a possibility in many cultures. Usually this choice has important legal
implications like accepting/refusing debts of the giver. In the blockchain
context it may translates in accepting/refusing the result of some smart 
contracts that are executing. This contributes to the inherent complexity of 
inheritance process within the blockchain context. If the smart contract wait
for an acceptance/refusal tick how does it warn the interested parties? What if
no one accepts? Etc. The list of questions grows larger as closer as you look
into the issues. Some problems are just programming issues but others are
specific to the inheritance process. The thing is that inheritance is a global
functionality of a community and not just a personal issue. Just like the
functionality of funds transfer is not up to the individuals. What is up to the
individual is the definition of the receiver and the amount to be transferred.
But the idea is not up to the whim of the user. Some parts of the inheritance
process are of this nature. To pinpoint which ones and to find out how they
could be implemented in a blockchain world remain open questions.  

   \subsection{Atomicity of wills}
   \label{subsec:atom_w}

If only because of technical limits, the inheritance process has always been,
historically, an atomic process in the sense that it was performed locally and
executed as a batch. It could take time to gather all heirs and
compute who inherits what but the process itself is clearly identified. It is
no longer the case with crypto-assets and it raises specific challenges that are
neither purely legal nor purely technical. For once, crypto-assets are not going
to rely on a single blockchain. Each blockchain may propose its version of
inheritance.

In \cite{Weyl2022} the case for a decentralized society is made. Among the many
points raised, the idea that the key primitives are the accounts, or the wallets,
is central:

\begin{quote}
Note there is no requirement for a Soul to be linked to a legal name, or for
there to be any protocol-level attempt to ensure “one Soul per human.” A Soul
could be a persistent pseudonym with a range of SBTs that cannot easily be
linked   
\end{quote}

The whole idea of inheritance is impacted by this. The questions raised are not
only of the technological realm.  
 
Another non trivial evolution, that has both societal and technological
implications, is that, by essence, a Blockchain is not an object that can be
located in a specific place. Therefore, it is not clear how the legal
liabilities may be inferred. Typically something like "which court has
jurisdiction?" is not an easy question to settle. The same questions arise
regarding the tax code to be considered. 

Many pages of similar problems could be filled. They are not trivial issues. As
mentioned earlier, the body of laws, practices and customs linked to legacy is
immense. It looks like that the useful abstractions in order to think about
those issues have yet to emerge. In that respect the inheritance problem is very
singular: the problem to solve is not yet clearly defined and cannot be possibly
well defined. Solutions and concepts are going to emerge from practice and
adoption. Yet those are not issues that can be totally outsourced, from a
blockchain point of view.    

\section{Conclusion}
\label{sec:conclusion}
In this paper we have discussed many problems and ideas around the theme of the
inheritance of crypto-assets. There are several levels to consider. They range
from a broad societal point of view down to the technical details and the 
specifics of protocols. Most of the challenges remain open at the time of 
writing. It is not yet clear either what should be part of the blockchain
infrastructure, and what should be delegated to the outside world. This question
is an integral part of the discussion that has to be done.   

Studies on the inheritance issue of crypto-assets are interesting  in and of
themselves. They have the potential to lead to interesting results in the field
of blockchains in general.  
 
\begin{itemize}
   \item The inheritance process is partly a social issue. As such every
technological proposition should be made with that caveat in mind. In 
particular, it means that solely technical solutions are not going to make it. 
   \item Any proposition should be such that it includes new possibilities
for the blockchain to interact with the outside world. Those interactions should
be more sophisticated than what the current implementation of Oracles offers. 
Typically, the fact that heirs may not necessarily be users of a  blockchain 
implies that there are tools that have to be developed in order to allow 
interactions from the blockchain towards the real world. Which is a dual problem 
than the traditional problem that is addressed by Oracles.
   \item By its very nature the inheritance process is a very long time problem.
The time horizon counts in decade. Very few issues have this property, but as
the move towards a digital society accelerate, more and more aspects of our
lives will be tackled using digital technologies. And so more and more digital
products will accompany us throughout our lives. How to manage such a kind of
products? How can they be tested? Are some of the questions that will have to be
answered.   
\end{itemize} 

The intrinsic social dimension of many aspects of the inheritance process
suggests that social media should be a useful tool to investigate. The
interesting feature of social media is that they are inherently distributed.
Moreover, they could be used to develop virtual identities that do not rely on a
number recorded in a database. Indeed, digital identity could be seen as the sum
of interactions within a social network rather than a number inside a
centralized database. This is of major importance because the whole point of
having a blockchain based mechanism for inheritance is to have a mechanism that
is not relying on third parties. How and with what level of warranties are the
important questions that future researches will have to address.      

%%%%%%%%%%%%%%%%%%%%%%%%%%%%%%%%%%%%%%%%%%%%%%%%%%%%%%%%%%%%%%%%%%%%%%%%%%%%%%%%
%%%%%%%%%%%%%%%%%%%%%%%%%%%%%%%%%%%%%%%%%%%%%%%%%%%%%%%%%%%%%%%%%%%%%%%%%%%%%%%%
  \bibliographystyle{alpha} 
  \bibliography{biblio}

\newcommand{\etalchar}[1]{$^{#1}$}
\begin{thebibliography}{ABRSS20}

\bibitem[ABRSS20]{Oracle}
Hamda Al-Breiki, Muhammad Habib~Ur Rehman, Khaled Salah, and Davor Svetinovic.
\newblock Trustworthy blockchain oracles: Review, comparison, and open research
  challenges.
\newblock {\em IEEE Access}, 8:85675--85685, 2020.

\bibitem[Bei11]{beimel2011}
Amos Beimel.
\newblock Secret-sharing schemes: A survey.
\newblock pages 11--46, 05 2011.

\bibitem[Bha22]{imf2022}
Gita Bhatt.
\newblock Reimagining money in the age of crypto and central bank digital
  currency.
\newblock {\em International Monetary Fund Blog}, September 2022.

\bibitem[Bib]{Bible}
{\em The Bible}.

\bibitem[But13]{Buterin2013}
Vitalik Buterin.
\newblock Ethereum white paper: A next generation smart contract \&
  decentralized application platform.
\newblock 2013.

\bibitem[Cas22]{Casa}
Casa,
  \href{https://keys.casa/bitcoin-inheritance-plan}{https://keys.casa/bitcoin-inheritance-plan},
  2022.

\bibitem[Ceb21]{Cebo2021}
Daniel Cebo.
\newblock Scientific relevance and future of digital immortality and virtual
  humans.
\newblock {\em CoRR}, abs/2101.06105, 2021.

\bibitem[DCB22]{Dickerson-Southworth22}
Graeme Dickerson{-}Southworth, Brian Chen, and James Braman.
\newblock Securing the accounts of the deceased: Implications of compromised
  profiles.
\newblock In Constantine Stephanidis, Margherita Antona, and Stavroula Ntoa,
  editors, {\em {HCI} International 2022 Posters - 24th International
  Conference on Human-Computer Interaction, {HCII} 2022, Virtual Event, June 26
  - July 1, 2022, Proceedings, Part {IV}}, volume 1583 of {\em Communications
  in Computer and Information Science}, pages 467--472. Springer, 2022.

\bibitem[HHL{\etalchar{+}}21]{smart2022}
Tharaka Hewa, Yining Hu, Madhusanka Liyanage, Salil Kanhare, and Mika
  Ylianttila.
\newblock Survey on blockchain-based smart contracts: Technical aspects and
  future research.
\newblock {\em IEEE Access}, 03 2021.

\bibitem[Hom]{Odyssey}
Homer.
\newblock {\em The Odyssey}.

\bibitem[Nak09]{Nakamoto_bitcoin}
Satoshi Nakamoto.
\newblock Bitcoin: A peer-to-peer electronic cash system,
  \href{http://bitcoin.org/bitcoin.pdf}{http://bitcoin.org/bitcoin.pdf}, 2009.

\bibitem[Pro22]{tales_from_crypt}
Frédéric Prost.
\newblock On the heritage of crypto assets -- tales from the crypt protocol,
  2022.

\bibitem[sar20]{sarcophagus}
Sarcophagus - a decentralized dead man switch,
  \href{https://sarcophagus.io/}{https://sarcophagus.io/}, 2020.

\bibitem[Sri22]{networkstate}
Balaji Srinivasan.
\newblock {\em The Network State: How To Start a New Country}.
\newblock 2022.

\bibitem[Ter22]{Ternoa}
Ternoa - white paper,
  \href{https://github.com/capsule-corp-ternoa/white-paper/blob/main/white-paper-en.md}
  {https://github.com/capsule-corp-ternoa/white-paper/blob/main/white-paper-en.md},
  2022.

\bibitem[WJ17]{Williams2017ArchainAO}
Sam~A. Williams and Will Jones.
\newblock Archain: An open, irrevocable, unforgeable and uncensorable archive
  for the internet.
\newblock 2017.

\bibitem[WOB22]{Weyl2022}
E.~Glen Weyl, Puja Ohlhaver, and Vitalik Buterin.
\newblock Decentralized society: Finding web3's soul.
\newblock {\em SSRN Electronic Journal}, 2022.

\end{thebibliography}

\end{document}